\documentclass{kluwer}
\usepackage{epsfig}
\begin{document}
\begin{article}

\begin{opening}
\title{Chemically Consistent Evolutionary Models with Dust}
\author{{\bf C.S. M\"oller}, U. Fritze--v.Alvensleben \& K.J. Fricke
         \email{cmoelle@uni-sw.gwdg.de}}
\institute{Universit{\"a}ts--Sternwarte G{\"o}ttingen, Germany}
\author{D. Calzetti}
\institute{Space Telescope Science Institute, USA}

\begin{abstract}

As a tool to interpret nearby and high redshift galaxy data from
optical to K-band we present our chemically consistent
spectrophotometric evolutionary synthesis models. 
These models take into account
the increasing initial metallicity of successive stellar generations
using recently published metallicity dependent stellar evolutionary
tracks, stellar yields and model atmosphere spectra. 

The influence of the metallicity is analysed. Dust absorption is
included on the basis of gas content and abundance as it varies with
time and galaxy type. 
We compare our models with IUE template spectra 
and are able to predict UV fluxes for different spectral types.
Combined with a cosmological model we
obtain evolutionary and k-corrections for various galaxy
types and show the differences to models using only solar metallicity input
physics as a function of redshift, wavelength band and galaxy type.

\end{abstract}
\end{opening}

\section{Introduction}

 The data on high and very high redshift galaxies is rapidly increasing 
these days. 
Many U- and B-dropout galaxies with photometric 
redshift estimates are known from the HST HDF, more than 550 have confirmed 
redshifts at z$\sim$3 (\citeauthor{Stei} 1998). 
Many more will be detected in ground-based large and
deep field projects (e.g. FORS -- Deep Field on VLT). 
Deep surveys are conducted at all 
wavelengths from UV through IR and far into the submm range (SCUBA on JCMT). 
These brilliant data require mature galaxy evolution models for adequate 
interpretation. Ideally, these models should cover 
the wavelength range
from UV through FIR and submm and describe the evolution of 
as many observable quantities as possible (spectrum, luminosities, colours, 
emission and absorption features for the stellar population, the gas content 
and a large number of element abundances for the ISM) and, at the same time,
be as simple as possible, involving the smallest possible number of 
free parameters.
A realistic galaxy 
evolution model should consistently take into account both the age and 
metallicity distributions of the stellar populations that enevitably
result from any extended SF history. 
This is what we attempt with our chemically consistent spectrophotometric, 
chemical and cosmological evolutionary synthesis model.
 The chemical evolution 
aspects of this model are presented in \inlinecite{Lin}) and in
Fritze-v. Alvensleben et al. (this conference)
in comparison with and interpretation of the observed redshift evolution of 
damped Ly$\alpha$ (DLA) abundances. 

Spectrophotometric and cosmological evolutionary synthesis models generally 
applied in current interpretations of high redshift galaxy data are using 
solar metallicity input physics (Bruzual \& Charlot 1993), 
Guiderdoni \& Rocca -- Vol\-mer\-ange 1987, 1988,  
Fioc \& Rocca -- Vol\-mer\-ange (1997), Poggianti (1997), Bressan et al. 1994)
together with 
specific parametrisations for the SF histories of various spectral types
of galaxies.
The first attempts to account for non-solar abundances and their impact on the 
photometric evolution of galaxies go back to Arimoto \& Yoshii (1986). 
Einsel et al. (1995) used more recent and complete stellar evolutionary 
tracks and colour calibrations for initial stellar metallcities 
$10^{-4} ...  4 \cdot 10^{-2}$ to describe the photometric evolution of galaxy 
types E through Sd. In M\"oller et al. (1997) we introduced the 
concept of 
chemical consistency into the spectrophotometric evolution of galaxies 
and investigated the time evolution of ISM metallicity and 
luminosity -- weighted mean stellar metallicities in various wavelength 
bands. For models that well agree with observed template spectra 
(Kennicutt 1992) of various types (E, Sb, Sd) we gave decompositions of 
the total light emitted at wavelengths from U through K in terms of 
luminosity contributions from various metallicity subpopulations. 
This clearly shows the considerable widths of the metallicity 
distributions in all 3 galaxy types. 

While in earlier investigations the main focus was on the role of 
${\rm Z > Z_{\odot}}$ stars in bulges and the centers of luminous elliptical 
galaxies, 
it is clear by today that the average stellar metallicity is 
${\rm \sim Z_{\odot}}$ in L$^{\ast}$ ellipticals
e.g. (\citeauthor{Loe} 1997)
and ${\rm < Z_{\odot}}$ 
in lower luminosity ellipticals  
as well as in bulges (Mc Williams \& Rich 1994). 
HII region abundances in spirals show negative radial gradients and 
characteristic oxygen abundances (at ${\rm r = 1 R_e}$) in the range 
${\rm 0.5\  ... \ 1.6 \cdot Z_{\odot}}$ (cf. \citeauthor{Oey} 1993, 
\citeauthor{Zar} 1994, \citeauthor{Fer} 1998). 
These characteristic HII region 
abundances are well reproduced by our chemically consistent chemical evolution 
models for galaxy types Sa through Sd and are 
expected to give an upper limit to the metallicity of the youngest stars. 
Galactic B-star abundances point to a modest
${\rm Z \sim \frac{1}{2} \cdot Z_{\odot}}$ in the ISM of the Milky Way 
(e.g. \citeauthor{Kil} 1997).
Thus, it is evident that sub-solar abundances 
as well as chemically consistent modelling
become more and more 
important for the global properties of 
galaxies when going to later spiral types or dwarf galaxies,
already at 
${\rm z \sim 0}$, and even more so towards higher redshift. 

In recent years it has also become increasingly clear that dust in galaxies
 plays a non-negligible role in determining the appearance and their
 observed spectral energy distributions both for nearby and high redshift
 galaxies. 
 Dust obscuration of a factor of 
 a few has been inferred in the spectral energy distributions of young
 galaxies at redshift z $\geq$ 1 (Steidel et al. 1999, Glazebrook et al.
 1999). 
In light of the potential importance of dust, 
we present our galaxy evolution models for both cases with/without dust.
 
 Here we present the extended spectrophotometric model which includes
 the effects of dust in a chemically consistent manner, 
 with the amount of dust tied to the amount and metallicity of gas as
 the evolve with time for various galaxy types.

\section{Model description}

Our galaxy evolution model was first described by
Fritze - v.A. (1989),
the extended version allowing for a chemically consistent 
modelling for the 
photometric evolution and in detail by 
\inlinecite{Moe2}, 
 for the spectral and 
spectrophotometric evolution.

In the following we briefly outline 
the principle of our concept of chemical consistency which we 
consider an important step towards a more realistic galaxy modelling. 
In contrast to single burst single metallicity stellar populations 
like star clusters 
our chemically consistent galaxy evolution model follows the 
metal enrichment of the ISM and accounts for the resulting 
metallicity distribution 
of the composite stellar population, both with respect to 
the evolution of ISM abundances (\citeauthor{Lin} 1998) and 
to the spectral evolution as presented here.
We use various sets of stellar tracks 
and yields from the Padova group (\citeauthor{Bre} 1993,
\citeauthor{Fag} 1994a,b,c) and from 
\inlinecite{Cha} for m$_{\ast} < 0.8 M_{\odot}$
for five different metallicities from
$\rm Z= 4 \cdot 10^{-4}$ to $5 \cdot 10^{-2}$.
The evolution of each star
is followed in the HR diagram from birth to its final phases
  for five discrete metallicity ranges. 
If the ISM metallicity increases above one of our
limiting metallicities, the evolution of stars formed thereafter is 
followed with the tracks for the next higher metallicity.
At any timestep the HRD population is used to synthesize an integrated 
galaxy spectrum from a library of stellar spectra. This library 
contains 
stellar model atmosphere spectra
from UV to the IR for all spectral types, luminosity classes and 
5 metallicities (\citeauthor{Lej} 1998). 
The total galaxy spectrum is obtained by summing the stellar spectra, 
weighted by the population of the HRD for each metallicity and, 
finally, by coadding the spectra of the 
various single metallicity subpopulations. 

For a given IMF, the spectral galaxy types 
E, ..., Sd are described by appropriate SF histories. 
Our E/S0 model has a 
SFR $\rm \sim e^{\frac{-t}{t_{\ast}}}$ with an e-folding time 
$\rm t_{\ast} = 1 - 1.5$ Gyr, while for spiral models  
we assume SFRs linearly proportional to the gas-to-total mass ratio  with
characteristic timescales for SF $\rm t_{\ast}$ = 4-5, 7-8, 9-10 Gyr for
the Sa, Sb, Sc model respectively. 
The Sd model is described by a constant SFR.
The total mass of the galaxy is $\rm 2 \cdot 10^{11} M_{\odot}$ (E/S0), 
$\rm 10^{11} M_{\odot}$ (Sa,...,Sc), $\rm 5 \cdot 10^{10} M_{\odot}$ (Sd)
to yield after a Hubble time the observed type specific M$_B^{\ast}$.

Our model is simply a closed box with instantaneous gas mixing,
but it fully takes
into account the finite stellar lifetimes.

We adopt a simplified model of the dust distribution in galaxies, 
discriminating between spheroidal and disk configurations. For purely
sphe\-roi\-dal systems (e.g. ellipticals), we assume that the dust has the
same distribution as the stars, and that the two components are well mixed.
Thus, dust extinguishes light according to the standard formula
$ \frac{1-e^{\tau}}\tau $, with $\tau=\tau(\lambda)$ the optical depth at 
wavelength $\lambda$. 
For disk systems, we assume the dust is distributed homogeneously in 
the disk, with a vertical scale height which is the same as that of the young,
UV-bright stellar component (Wang \& Heckman 1996). 
The ratio of dust-to-stellar scale heights decreases from unity in the UV to
$\sim $ 0.25 at optical wavelengths where 
the emission is dominated by
older stars. This configuration mimics the 
disk/bulge separation as observed (e.g. 
Kylafis \& Bahcall 1987). The extinction curve we use 
is the one appopriate for the Small Magellanic Cloud (Bouchet et al.
1985). The choice of the extinction curve has very little, if any,
effect longward of 3000 \AA , where all three known extinction curves
(MW, LMC, SMC) are similar; in the UV, however, differences
are more important (cf. Figure 1 in Calzetti et al. 1994).
Since we are modelling galaxies with a wide variety of properties, we
adopt as a preliminary approach the SMC extinction curve in the 
UV. This is equivalent to picking up the low-metallicity, moderately
star-forming environment of the SMC as typical for
the galaxies we model.
We calculate the extinction in a chemically consistent way assuming
that the amount of dust is proportional to the gas column density and the 
ISM metallicity which keeps dust/metals constant over 
the time evolution (Dwek 1998).

Combining the spectrophotometric time evolution with a cosmological 
model and some assumed redshift of galaxy formation we calculate the 
evolutionary and cosmological corrections as well as the evolution of 
apparent magnitudes from optical to NIR for various spectral types and
cosmological parameters taking into account the attenuation by 
intervening HI as described by \inlinecite{Mad}. 

\subsection{Extinction in different Hubble types}

In Figure 1 we show for various galaxy models the time evolution 
of ISM metallicity, gas content and extinction. In the following
we classify the various spectral types from E, S0,..., Sd with their
characteristic timescales.
After a Hubble time,  
our models not only have the observed colours and spectral
energy distributions 
of the respective nearby galaxy types
(see comparison with Kennicutt templates in
M\"oller et al. 1999) but also the observed average
ISM abundances at $\sim 1 R_e$, gas content and E(B-V) of nearby galaxies.
To represent the wide range of observed 
properties for elliptical and S0's we calculate two different models 
varying $\rm t_{\ast}$ from 1 (a) to 1.5 (b) Gyr 
which has no remarkable effect on the spectral energy distribution 
except for 
the resulting extinction E(B-V) over the entire time evolution. 
After a Hubble time both E/S0 models have the same ISM metallicity
of about $Z\sim Z_{\odot}$. While in model a) SF stops
after 2 Gyr and the amount of gas increases to a few percent due to the 
stellar yields, the star formation in model b) continues until the gas 
content is negligible. Therefore the extinction in both models varies between
E(B-V) = 0 and 0.1 after a Hubble time.
These values are in good agreement with observations of field
ellipticals (Goufrooij et al. 1994), for E(B-V) can go up to 0.2.
The influence of the IMF is seen in the comparison 
of the two Sa models with Salpeter IMF
(c) and Scalo IMF (d). 
All other models are calculated with Salpeter IMF.
The model with Salpeter
IMF produces more metals and therefore results in a higher metallicity but also
consumes more gas for the same SFR and gives a lower extinction after a Hubble time.
The model with $\rm t_{\ast} = 8$ Gyr (e) shows the highest 
present day extinction of about 
0.3 to 0.4 mag which is also observed in \inlinecite{Gon}. 
The chemically consistent model with constant SF also shows an
evolutionary effect because of the increasing abundances and decreasing gas
content not only in the time evolution of its SED 
but also due to its increasing extinction. 

In all models, except for the constant SFR, we see that the evolution of the 
extinction over redshift
shows a maximum at high z, which is different for the various spectral
types, and then decreases again to very high z due to the low metallicity.

 \begin{figure}
 \centerline{\epsfig{file=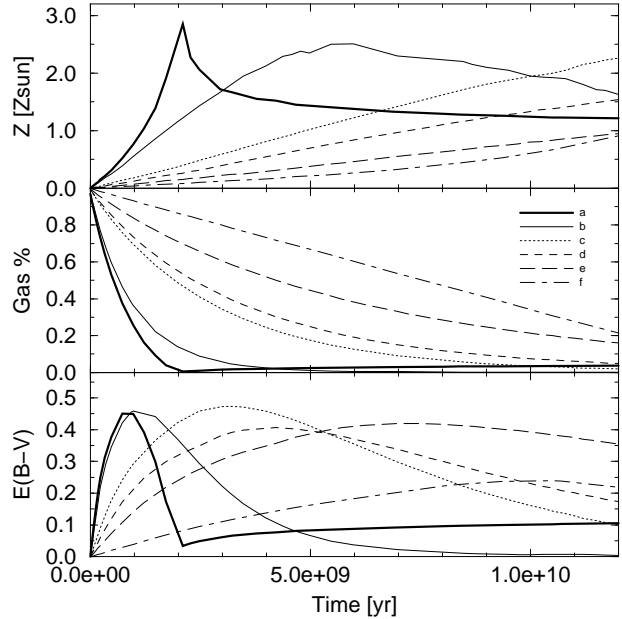, width=11cm, angle=0}}
 \caption{Time evolution of ISM metallicity (top), 
          gas content (middle) and extinc\-tion (bottom) for galaxy models
          E/S0 (a) and (b), Sa (b) and (c), Sb (e), Sd (f).}
 \end{figure}

\section{Comparision with nearby galaxies}

Our model describes the entire spectrophotometric evolution of field
galaxies, e.g. metallicities and colours measured at $\rm 1~ R_{eff}$.
In particular our E model represents a
medium luminosity elliptical galaxy. We have
shown that our model well describes the stellar metallicity observed by 
absorption indices (\citeauthor{Moe2} 1997) and the chemical evolution
of nearby and high redshift spiral galaxies (\citeauthor{Lin} 1999).
Our model SEDs for various types E, Sa,..., Sd show very good agreement 
with the templates NGC 3379, NGC 3368, ..., NGC 4449  from
Kennicutt's atlas in the wavelength range (3600 - 6800) \AA .

The galaxy NGC 1553 (Kinney et al. 1996) is best reproduced by a 
$\rm t_{\ast} = 1$ Gyr 
model with an $\sim$ 11 Gyr old stellar population, a mean ISM metallicity
of Z= 1.2 $Z_{\odot}$ and E(B-V) = 0.1.

The UV and optical spectrum of NGC 210, a template Sb galaxy, from
Kinney et al.'s atlas are compared to our model spectra in Figure 2. 
While our Sb or ($t_{\ast}$ = 8 Gyr) model with or without dust only gives
a very poor match, we find good agreement 
over the entire wavelength range (912 - 10000 \AA )
with our $t_{\ast}$ = 1 Gyr model at an
age of the stellar population of $\sim $ 9 Gyr, $Z=1.3 Z_{\odot}$ and E(B-V)=0.1.
The point is that for nearby galaxies the IUE Aperture of 10"x20" only covers the
central 300 - 500 pc which are dominated by the bulge component. 
The optical aperture in this case matches the one in the UV so that the
agreement with our $t_{\ast}$ = 1 Gyr model, which is appropriate for speroidals
and bulge components, is a result of this small aperture effect: while the
integrated spectrum of the Sb template NGC 210 should be well described by 
our $t_{\ast}$ = 7-8 Gyr model as is e.g. the case for Kennicutt's integrated
Sb template spectrum of NGC 3147, its bulge spectrum clearly requires a short
SF timescale $t_{\ast}$ = 1 Gyr.

It would be necessary to observe the UV flux over a wider area.
So far our models give predictions for the 
extinction and the metallicity, 
and it should be possible to disentangle extinction and metallicity
with a wide wavelength basis from
far UV to opt. or NIR bands.
 
 \begin{figure}
 \centerline{\epsfig{file=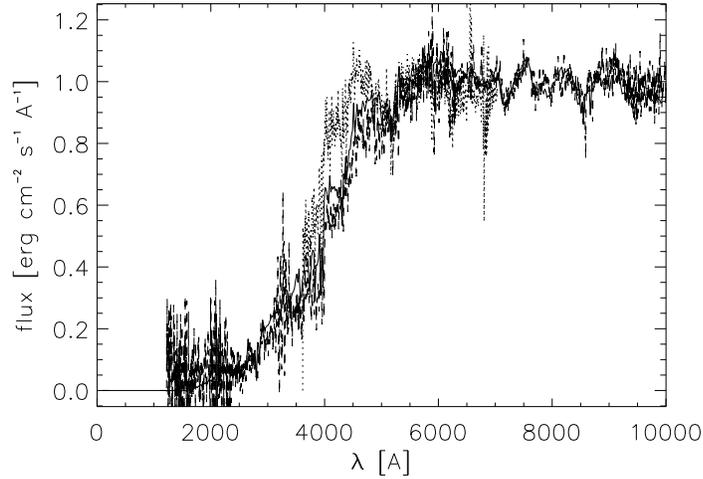, width=10cm, angle=0}}
 \caption{Comparision of Sb templates NGC3147 (dotted), NGC210 (dashed) with
          galaxy model $t_{\ast}$ = 1 Gyr (solid). Fluxes are 
          normalized at 5500 \AA .}
 \end{figure}
\section{Summary and outlook}

We have extented our spectrophotometric and chemical evolutionary
synthesis model in a chemically consistent
way. 
We present the time and redshift evolution of the extinction in various
galaxy types resulting from the evolution of their gas contents and 
metallicities.
Comparing our model SED's with templates from Kennicutt's and
Kinney et al.'s atlas we show the detailed agreement of our model spectra
with {\bf integrated} spectra of galaxies and point out the
importance of aperture effects on the example of an Sb galaxy. 
 
  We have compiled a large grid of evolutionary and cosmological
corrections from UV to IR and compare the models using only solar 
metallicity with the chemically consistent models (M\"oller et al. 1998).
A detailed description and comparison of model results w/o dust 
with observations 
of colors and  luminosity of high redshift galaxies will be presented
in M\"oller et al. 1999.


\end{article}

\end{document}